\documentclass[psfig,times]{aa}  
\input{psfig.sty}
\input{times.sty}
\usepackage{times}           
\begin{document}
\def\Mr{${\rm M}_{\rm R}$ }
\def\mr{${\rm m}_{\rm R}$ }
\def\re{${\rm r}_{\rm e}$ }
\def\1es{1ES 1741+196}
\thesaurus{  03                     
             (03.13.2;              
             11.01.2;               
             11.02.2: 1ES 1741+196; 
             11.09.2;               
             11.16.1)}              
\title{1ES 1741+196: a BL Lacertae object in a triplet of interacting 
galaxies?
\thanks{Based on observations made with the Nordic Optical
Telescope, operated on the island of La Palma, jointly by Denmark, Finland,
Iceland, Norway, and Sweden, in the Spanish Observatorio del Roque
de los Muchachos of the Institute de Astrofisica de Canarias.}
$^,$
\thanks{ Based
on observations collected at the German-Spanish Astronomical Centre,
Calar Alto, operated by the Max-Planck-Institut f\"ur Astronomie,
Heidelberg, jointly with the Spanish National Commission for Astronomy}
}
\author{J. Heidt\inst{1}, K. Nilsson\inst{2}, J.W. Fried\inst{3}, 
L.O. Takalo\inst{2}, and A. Sillanp\"a\"a\inst{2} }
\institute{Landessternwarte Heidelberg, K\"onigstuhl,
D$-$69117 Heidelberg, Germany
\and 
Tuorla Observatory, FIN$-$21500 Piikki\"o, Finland
\and
Max-Planck-Institut f\"ur Astronomie, K\"onigstuhl 17, D$-$69117 Heidelberg,
Germany}
\offprints{\ \protect\\
 J.~Heidt,~E$-$mail:~jheidt@lsw.uni$-$heidelberg.de}
\date{Received 17 February 1999 / Accepted 11 May 1999 }
\maketitle
\markboth{J. Heidt et al.: 1ES 1741+196: an interacting BL Lacertae object?}{}
\begin{abstract}

We present subarcsecond resolution imaging and spectroscopy of the 
BL Lac object \1es and neighboring galaxies. Based on 
2$-$dimensional modelling, the host galaxy of \1es is a very bright and 
large elliptical galaxy
(\Mr = $-$24.85, \re = 51.2 kpc) whose overall luminosity distribution  
deviates significantly from a de Vaucouleurs profile. 
It is one of the most luminous and largest BL Lac host galaxies known.

Closeby to \1es we found two companion galaxies at the same redshift
as the BL Lac itself. They are at projected distances of 7.2 and 25.2 kpc,
respectively. The closer companion galaxy can be best modelled by a Sa-type
galaxy, whereas the more distant companion galaxy is an elliptical. 
This is supported by their spectra. We detected a tidal tail emerging
from the closer companion galaxy which is possibly connected with the
more distant galaxy. Its surface brightness 
increases towards the closer companion galaxy, which suggests that material
has been released from that galaxy due to tidal forces.  

The flat luminosity profile ($\beta$ = 0.15),  high ellipticity 
($\epsilon$ = 0.35) of the host galaxy of \1es as well as its position angle 
along the impact parameter to the neighboring galaxies can be the result 
of tidal interaction. \1es may be a BL Lac object in a triplet 
of interacting galaxies.

\keywords{ Methods: data analysis -- Galaxies: active -- 
BL Lacertae objects: individual: 1ES 1741+196 -- Galaxies: interactions 
-- Galaxies: photometry}

\end{abstract}

\section{Introduction}

Studies of the host galaxies and close environment of BL Lac objects provide 
important insights into the mechanism, which could be responsible for
the extreme properties of these extragalactic objects. Following the standard
model, the mass accreting Black Hole in the center of AGN host galaxies
accounts for the extraordinary energy output of these objects.
Hence, the fuelling mechanism of the central engine and the role played by
tidal interaction with neighboring galaxies has to be clarified.
Due to the loss of angular momentum by tidal forces either material 
from a gas-rich neighbor or the gas in the host galaxy itself could
fuel the nucleus. Evidence for such a scenario have been claimed e.g. by
Hutchings \& Neff (1992).

While the nature of BL Lac host galaxies (at least at low-redshifts) 
is well understood (see e.g. Heidt 1999 for a recent review),
the immediate environment ($<$ 50 kpc) is as yet poorly studied.
This is mainly due to the faintness of most neighboring galaxies 
(3$-$5 mag fainter than the BL Lac) as well as the 
limited resolution available from ground. However, in recent years,
a noticeable number of BL Lac objects with close companion galaxies 
(e.g. Falomo et al. (1990, 1991, 1993), Falomo (1996), 
Heidt et al. (1999)) or signs of 
interaction have been observed (e.g. Falomo et al. 1995, Heidt et al. 1999).
This can be taken as evidence that tidal interaction is potentially important 
to the BL Lac phenomenon at least in these sources.

\1es (z = 0.083) is a member of the Einstein Slew Survey sample
of BL Lac objects (Perlman et al. 1996), whose 
host and environment has not been studied so far.
Therefore we carried out an an imaging and spectroscopical study the 
results of which
are presented here. Throughout the paper ${\rm H}_{\rm 0} =$ 50\ km\ 
${\rm s}^{\rm -1}\ {\rm Mpc}^{\rm -1}$ 
and ${\rm q}_{\rm 0} = 0$ is assumed.

\begin{figure*}
\centerline{\hbox{
\psfig{figure=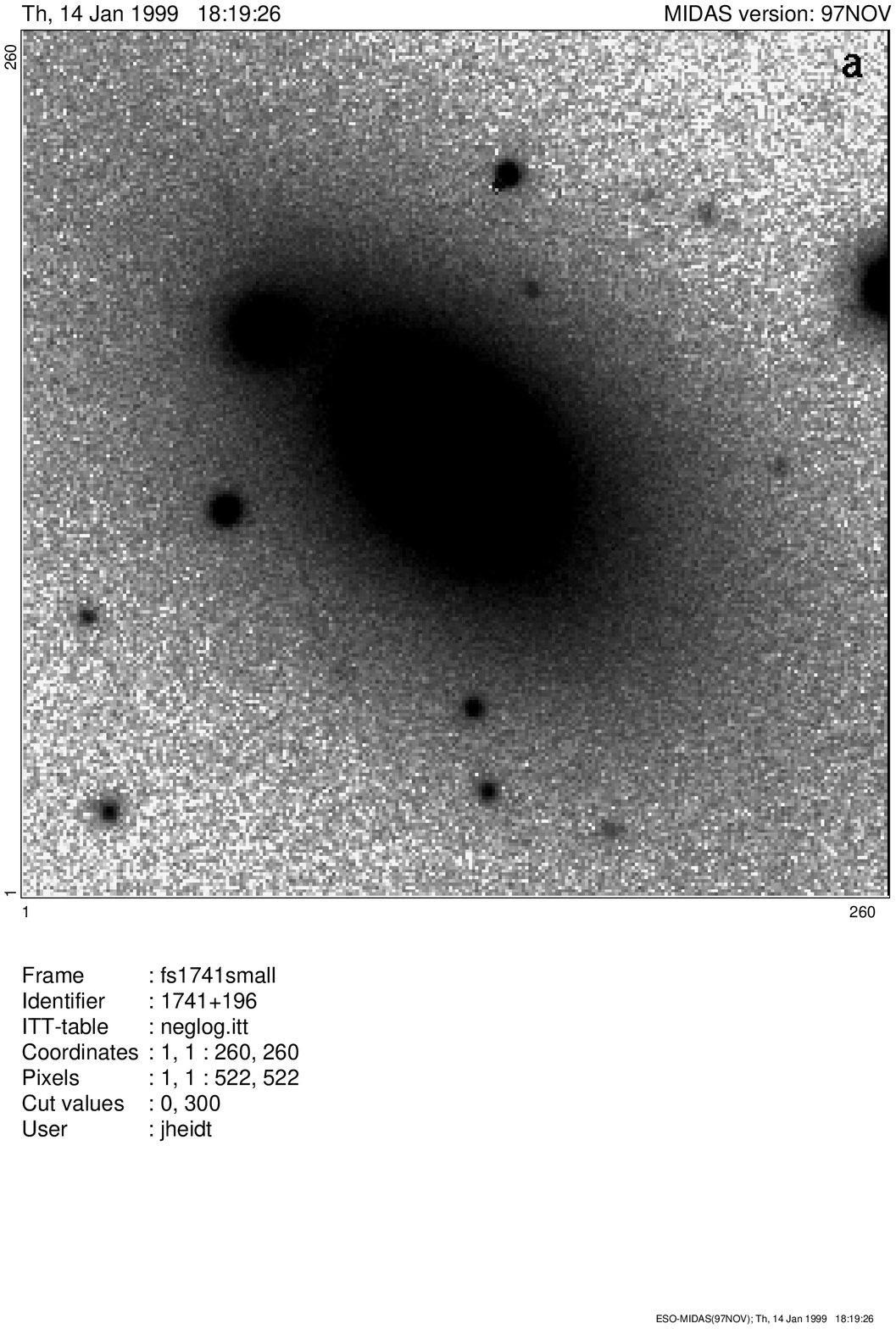,width=6cm,height=6cm,clip=t}
\hspace*{.5cm}
\psfig{figure=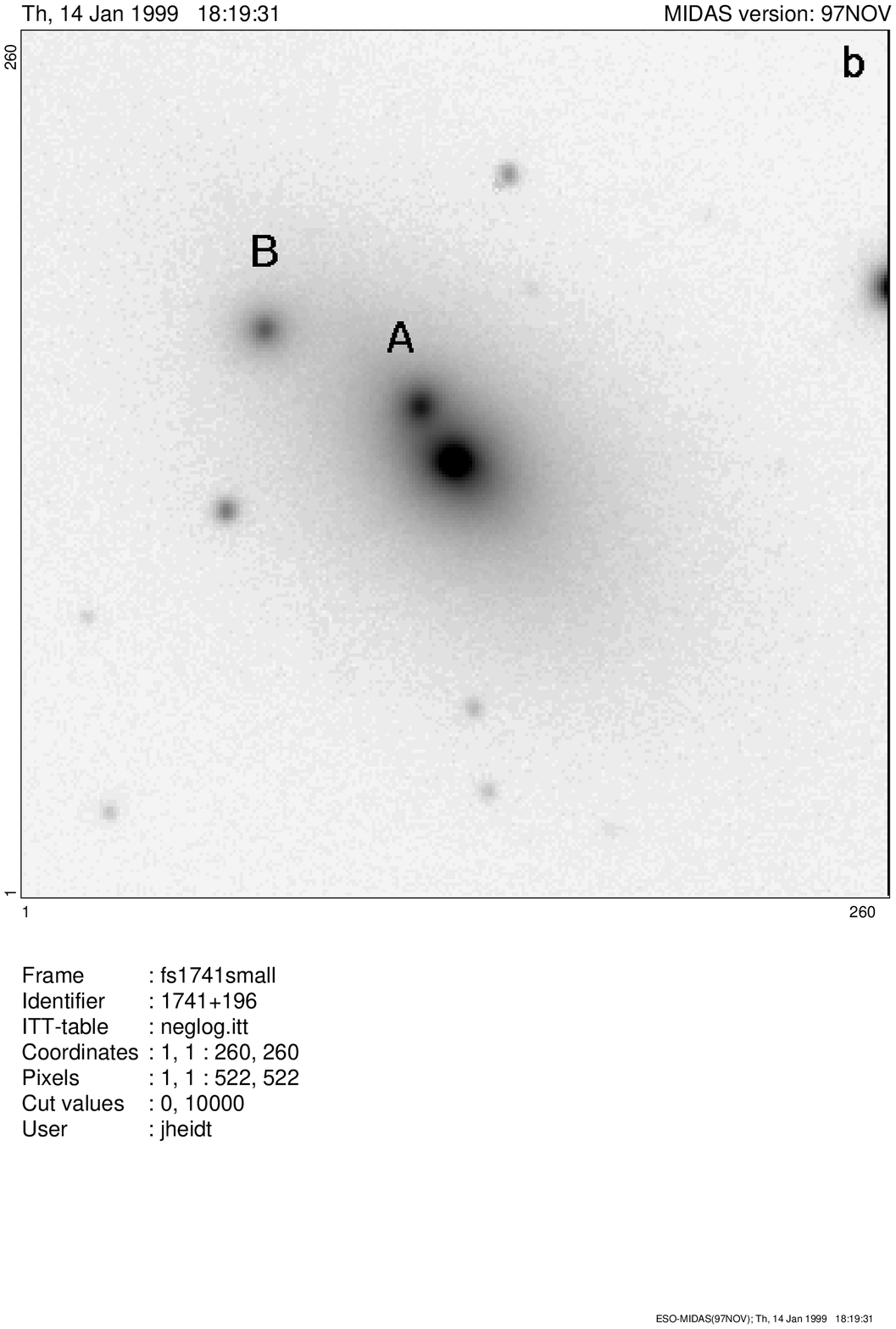,height=6cm,width=6cm,clip=t}
}}
\vspace*{.5cm}
\centerline{\hbox{
\psfig{figure=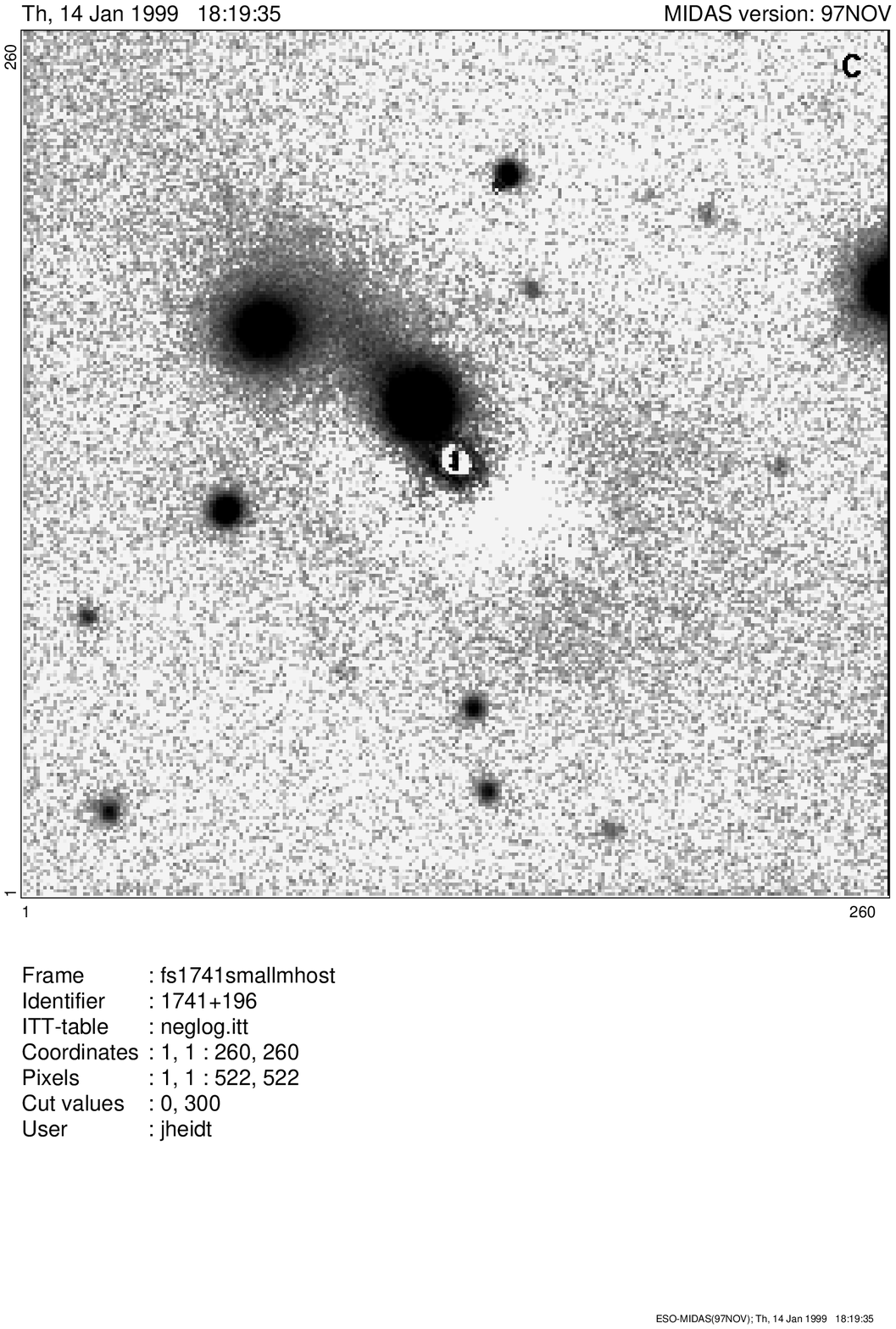,width=6cm,height=6cm,clip=t}
\hspace*{.5cm}
\psfig{figure=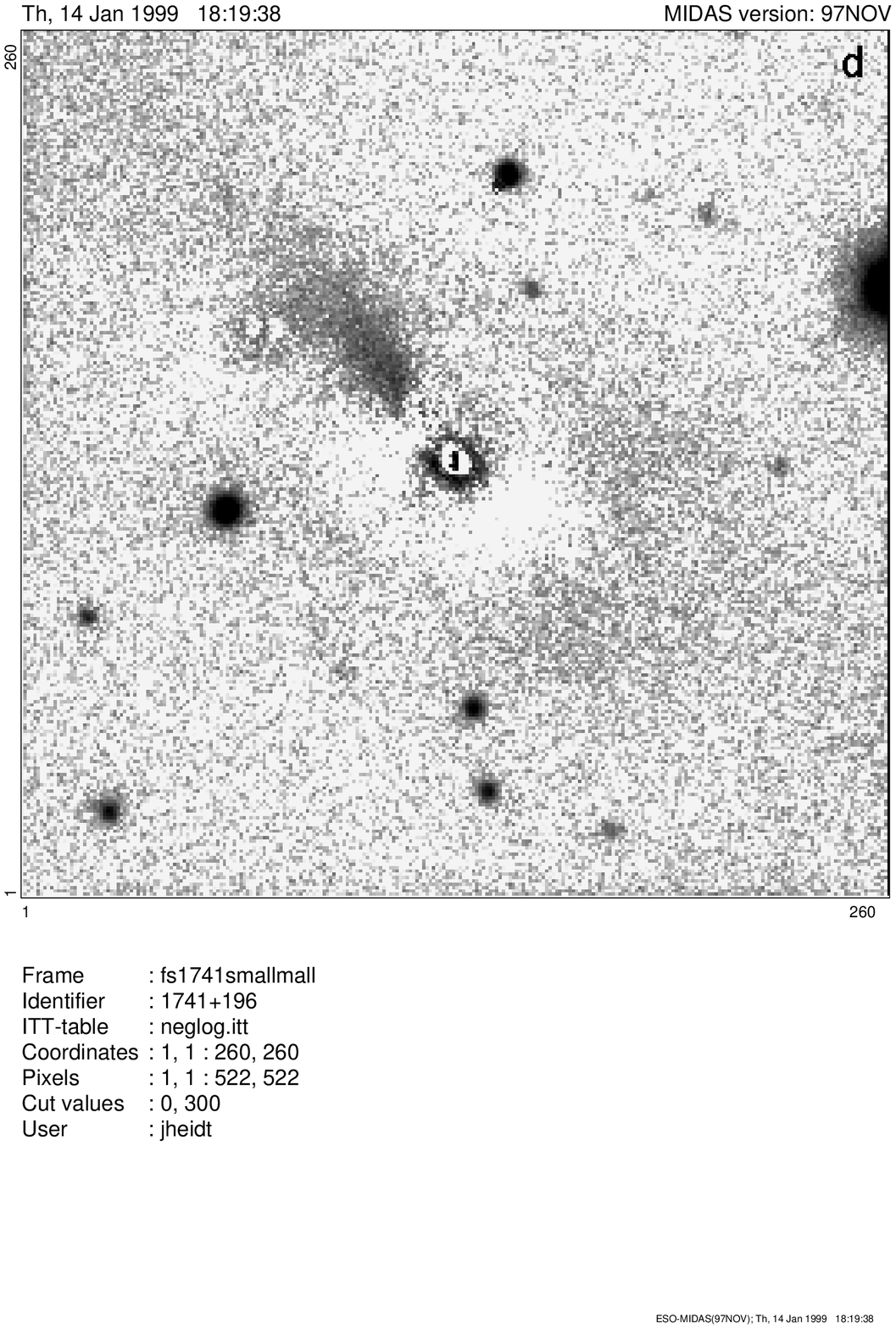,height=6cm,width=6cm,clip=t}
}}
\caption [] {a) Image of \1es in order to show the 
full extent of the host galaxy. 
Field is 46\arcsec$\times$46\arcsec\ 
(100$\times$100 kpc at z = 0.083). North is up, east to the left.
b) Same as a) but with different dynamic 
range to show the two nearby companion galaxies. They are labelled ``A'' 
and ``B''. 
c) Same image after subtraction of the model for the host galaxy
of 1ES 1741+196 and the core. Note the tidal tail emerging from galaxy A
possibly connected to galaxy B.
d) Same as c) with the models for the two companion galaxies
subtracted to show the tidal tail more clearly.
}
\end{figure*}

\section{Observations and data reduction}

High-resolution imaging data of \1es were taken with the Nordic Optical 
Telescope on the night July 12/13 1996. A 1k CCD (scale $0.176\arcsec$/pixel)
and an R filter was used. We observed the BL Lac for 840 sec in total,
split in several exposures to avoid saturation of the BL Lac.
The night was photometric, standard stars from Landolt (1983) were 
frequently observed to set the zero point.
The data were reduced (debiased, flatfielded using twilight flatfields),
cleaned of cosmic ray tracks, aligned and coadded. The FWHM on the final 
coadded frame is $0.78\arcsec$.

A longslit spectrum of \1es and two nearby companion galaxies was taken with 
the Calar Alto 3.5m telescope on the night April 21/22 1998. The focal reducer
MOSCA with a 2k CCD (scale $0.32\arcsec$/pixel) and grism GREEN\_500 
($\lambda\lambda \sim 4000 - 8000$ \AA\ with 1.9 \AA/pixel) was used. 
The instrumental resolution with the $2\arcsec$ slit was 14 \AA. The 
slit was oriented at PA = $57^{\circ}$ to cover the centers of the two nearby 
companion galaxies and the host galaxy of \1es. Two spectra of 1800 sec 
each were taken.
They were bias-subtracted, flatfielded, corrected for night sky background and
averaged. Wavelength calibration was carried out using HgAr calibration lamp
exposures. Flux calibration was derived from observations
of the standard star BD+$33^{\circ}$2642 (Oke 1990). 

\section{Data analysis and results}

Figs. 1a and b display \1es and its close environment with two dynamical 
ranges
to show the extent of the host galaxy (a) and the two nearby 
companion galaxies (b). The two nearby companion galaxies 
(labelled ``A'' and ``B'' in Figure 1b) are at projected distances of 
$3.3\arcsec$ and $12\arcsec$, respectively. 

In order to model the BL Lac and its host as well as the two companion
galaxies we applied a fully 2$-$dimensional fitting procedure 
to the images (for details see Heidt et al. 1999).
Before fitting companion galaxies A and B the uneven background produced
by the host galaxy of \1es has to be removed.
To achieve this, we used the ellipse fitting task ELLIPSE
in IRAF to model the core and host of \1es. This model was subtracted from
the image resulting in a flat background. After masking projected
stars etc. the two companion galaxies were fitted. Next we subtracted the
models for the companion galaxies from the original image, masked the central
regions of the subtracted companions (which show some residuals) and fitted
then the host and core of \1es. To verify the results of our fitting
procedure, we subtracted the models for the companion galaxies from 
the original image, and repeated the procedure above (i.e. made a model
of \1es with ELLIPSE, subtracted the model etc.). The results 
for the fit parameters between iteration 1 and 2 differ less than 1\%.

\begin{table}[t]
\caption[]{Results of our fits to the host of \1es and the two companion 
galaxies.}
\begin{tabular}{llcccrcrc}
\hline
 & & & & & \\
Object & $\beta$ & \mr  & \Mr & \re [\arcsec] & \re [kpc] \\
 & & & & &  \\
\hline
 & & & & &  \\
1ES 1741+196     & 0.15 & 14.07 & $-$24.85 & 23.4 & 51.2 \\
                 & 0.25 & 14.47 & $-$24.45 & 11.6 & 25.4 \\
Galaxy A (bulge) & 0.25 & 18.37 & $-$20.55 & 0.56 & 1.2  \\
Galaxy A (disk)  & 1    & 18.27 & $-$20.65 & 0.81 & 1.8  \\
Galaxy B         & 0.25 & 18.04 & $-$20.88 & 1.6  & 3.5  \\

 & & & & &  \\
\hline
\end{tabular}
\end{table}

We used seven different models (three galaxy models with and without a 
nuclear point source and a bulge+disk model) for the analysis. 
They were described by a 
generalized surface brightness
distribution with a shape parameter $\beta$ (Caon et al. 1993). We have chosen
$\beta$ = 1 (disk galaxy), $\beta$ = 0.25 (de Vaucouleurs) and
$\beta$ as free parameter. All galaxy models were convolved with the 
observed PSF, which was obtained by averaging several well exposed
stars on the frame. For the nuclear point source a scaled PSF was used.
PSF variations were proven to be negligible except in the central
regions ($\leq$ 2 pixel). Absolute magnitudes of the galaxies were 
calculated using K$-$corrections adopted from Bruzual (1983), the galactic 
extinction was estimated from Burstein \& Heiles (1982). 

The results of our fits are summarized in Table 1. The best fit for the host
of \1es was obtained with an elliptical galaxy model with $\beta$ = 0.15. 
The host is very bright (\Mr = $-$24.85), extremely large 
(\re = 51.2 kpc) and has a relatively high ellipticity ($\epsilon$ = 0.35) 
with position angle PA = $48^{\circ}$. The fit with 
the de Vaucouleurs model ($\beta$ = 0.25) was less satisfactory
($\chi^2_{\rm red}$ = 1.76 vs. 1.32),
but even then the host is very bright and large (\Mr = $-$ 24.45,
\re = 25.4 kpc). The decentering (host versus core centroids) is negligible
for both models ($\leq$ 0.04\arcsec). For comparison we give the 
results of both fits in Table 1. We determined the significance of the best fit
with $\beta$ = 0.15 by numerical simulations (see Heidt et al. 1999 for a 
description of the procedure). The result for $\beta$ = 0.15 is statistically
significant at a 10 $\sigma$ level. 

\begin{figure}
\psfig{figure=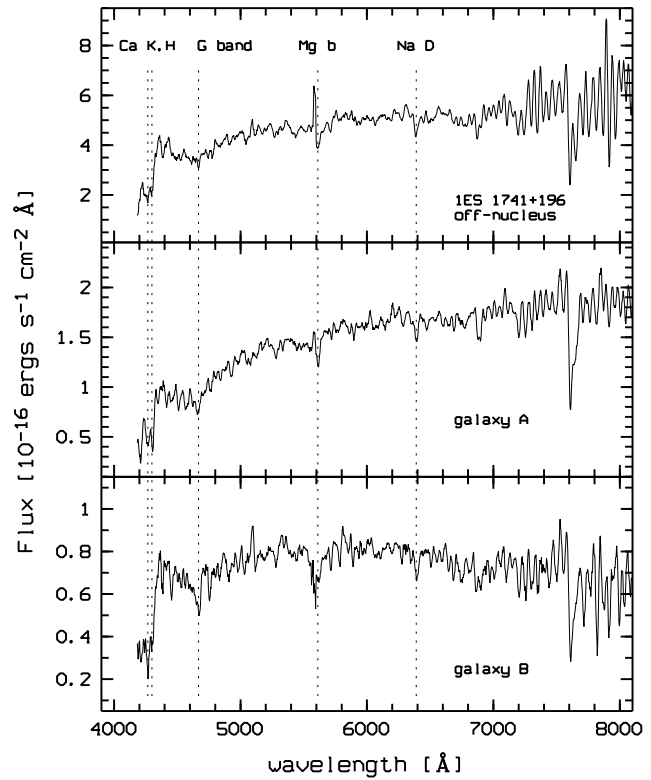,width=8.5cm,clip=t}
\caption [] {Spectra of the host of \1es and the companion galaxies A and B.
The identified absorption lines are marked.}
\end{figure}

Galaxy A can best be fitted by a system consisting of a disk and a bulge
(\mr (bulge) = 18.37 and \re (bulge) = 1.2 kpc, \mr (disk) = 18.27 and 
\re (disk) = 1.8 kpc), which is typical for a Sa-type galaxy.
The best fit for galaxy B was obtained with a de Vaucouleurs model
(\mr = 18.04, \re = 3.5 kpc). 

In Fig. 1c we show the image after subtraction for the model of \1es.
The two companion galaxies A and B and a low surface brightness feature 
between the two companion galaxies are clearly visible.
In order to examine the nature of the low surface brightness feature,
we further 
subtracted our models for the two companion galaxies from the image. 
The result is displayed in Fig. 1d. 
The feature is still present, the surface brightness at peak
is 24.5 mag/sq. arcsec. Its appearance  
is suggestive of a tidal tail,
indicating current interaction between both companion galaxies. It is 
more condensed towards companion galaxy A, which might 
imply that material is being stripped from that galaxy.
We emphasize that the tidal tail is a real feature and not an artifact
caused by our modelling procedure. It always
shows up irrespective of the models used (either IRAF/Ellipse task or our 
own procedure which both produce smooth models). 

Since the spectra of the two companion galaxies 
are contaminated by the contribution from the BL Lac host, 
we adopted a decomposition procedure to extract the 1-dimensional 
spectrum for each of the galaxies. 
Perpendicular to the dispersion, we fitted column by column
three Gaussians representing the contribution from the host of \1es
and the companion galaxies to the flux distribution by a least-squares method. 
The integrated flux of each Gaussian for each column was then used to 
derive the 1-dimensional spectra. This gave us confidence that the 
lines found in the spectra of the companion galaxies are not caused by the 
host of \1es. The spectra are shown in Fig. 2.

All three spectra show Ca K+H, G-band, Mg b and Na D in absorption typical 
for bulge-dominated  galaxies. This is consistent with the results of our 
fits to the images. No emission lines were detected. From the 
absorption lines we derive z = 0.084$\pm$0.001 for \1es and companion 
galaxy A and z = 0.085$\pm$0.002 for companion galaxy B. All three
galaxies are at the same redshift within the errors (actually the redshifts
differ by z = 0.0005). Our redshift for \1es is in accordance with the value
(z = 0.083) given by Perlman et al. (1996), who measured the same absorption
lines.

\section{Discussion}

With \Mr = $-$ 24.85 and \re = 51.2 kpc, the host of \1es is one of the 
brightest and largest BL Lac hosts known to date. This is true
even when the de Vaucouleurs model is used.
Then \Mr = $-$24.45 and \re = 25.4 kpc, which is still considerably brighter 
and larger than the typical BL Lac host (\Mr = $-$23.5 and \re = 10 kpc, 
Heidt 1999).
Simi\-lar half-light radii  have been found e.g. for PKS 0301-243 and 
PKS 0548-322 in R-band (Falomo 1995) and  H0414+009, MS 0419+197 and  
and PKS 1749+096 in r-band (Wurtz et al. 1996). However, only PKS 0548-322 
is of similar brightness (Falomo 1995). It is remarkable that the results for
PKS 0548-322 obtained by Falomo (1995) and Wurtz et al. (1996) differ
considerably (\re = 51 kpc versus 13.77 kpc and \Mr = $-$ 24.2 versus
$-$23.25, respectively).

According to our simulations, the deviation of the galaxy profile 
of the host of \1es from a
de Vaucouleurs profile is significant. A $\beta$ of 0.15 represents
flatter light distribution than $\beta$ = 0.25. This can 
be explained by tidal interaction with the two neighboring galaxies,
which a) have the same redshift as \1es and b) are at projected
distances of 7.2 kpc (galaxy A) and  26.3 kpc (galaxy B), respectively
and are thus within the half-light radius of the host of \1es.
During an encounter of galaxies, initial orbital energy of the galaxies is
transferred into internal energy, which in turn perturbes the initial
mass distribution. The galaxies expand along their impact parameter and
contract perpendicular to their impact parameter. Finally, the galaxies blow 
up and their luminosity profiles become flatter (Madejski \& Bien, 1993). 
The results of our fits to the host of \1es are consistent with this
scenario. The luminosity profile is flat, the galaxy is rather elliptical
and the PA = $48^{\circ}$ is approximately along the impact parameter between
\1es and the companion galaxies. This effect is not pronounced for the 
two companion galaxies, but here the situation is complicated due to the 
interaction by the galaxies themselves.

An interesting  observation is the tidal tail emerging from galaxy A
possibly connected to galaxy B. 
It is more condensed towards galaxy A,
which would suggest that material has been released from this galaxy.
Since galaxy A is most likely a bulge-dominated disk system, the material
could well be a mixture of stars and gas. Unfortunately, no emission lines,
which would be expected for Sa-type systems or which could be signs of 
recent star formation induced by tidal interaction can be found in the spectra.
This is not unexpected, however. First, the slit had a width of 2\arcsec\
thus probing the inner part of the galaxy dominated by the bulge. Secondly,
the slit orientation covered the tidal tail only in part. Finally, the 
whole system is polluted by the host galaxy of \1es, which makes it very hard 
to detect emission lines unless they are very strong. 

The observations of \1es presented here and the observations of 1ES 1440+122
and 1ES 1853+671 (Heidt et al. 1999) may offer an unique opportunity to
study nuclear activity induced by tidal forces in BL Lac objects.
All objects have relatively bright companion(s) within 10 kpc projected 
distance. Whereas 1ES 1440+122 has a very bright companion at the
same redshift (M. Dietrich, priv. com.) perhaps approaching
the BL Lac, \1es seems to be in an ongoing state of interaction with
its companion galaxies and 1ES 1853+671 has a companion which seems to be 
merging  with 1ES 1853+671 itself. Thus these three objects may form a 
homogeneous sequence from an early to a late stage of interaction. 
Unfortunately, the redshift of the companion galaxy to 1ES 1853+671 
is not known, which makes this consideration a bit speculative. 

One might ask, if these three BL Lac objects are typical for their class.
Close companion galaxies have often been observed  (e.g. Falomo 1990,
1996, Heidt et al. 1999), but in most cases they are relatively faint and
redshifts are unknown. As such the three 1ES BL Lac objects
are not untypical except the brightness of their companion galaxies.

A major drawback of this consideration is a clear demonstration that 
activity in BL Lac objects is triggered or maintained by gravitational
interaction. The discussion on this subject and its relevance to
AGN has a long and contrary discussed history. In one of the last papers 
dealing with this issue, De Robertis et al. (1998) compared the environments 
of a well defined sample each of Seyfert and "normal" galaxies
and found essentially no difference. Such a comparison has not been
conducted yet for any kind of radio-loud AGN. As already discussed in
Heidt et al. (1999) this is a tricky work, but urgently needed.

\acknowledgements{We thank the referee (Dr. J. Stocke)
for his critical comments. This work was supported by the DFG
(Sonderforschungsbereich 328) and the Finnish Academy of Sciences.}

\end{document}